\documentclass[useAMS,usenatbib,asm]{mn2e}
 \usepackage{graphicx,fleqn,times}

\title[Blue Stragglers in Arp~2]{An analysis of the blue straggler population in
the Sgr dSph globular cluster Arp~2\thanks{Based on observations carried out at ESO La 
Silla under program 081.C-0087(A).}}

\author[G. Carraro  and A.F. Seleznev]{Giovanni Carraro$^{1}$\thanks{On 
leave from Dipartimento di Astronomia, 
Universit\'a di Padova, Italy}\thanks{E-mail:
gcarraro@eso.org (GC); anton.seleznev@usu.ru (AFS)}
and Anton F. Seleznev$^{2}$\footnotemark[2]\\
$^{1}$European Southern Observatory, Alonso de Cordova 3107, 
Casilla 19001, Santiago 19, Chile\\
$^{2}$Astronomical Observatory, Ural State University, Lenin avenue. 51, Ekaterinburg 620083, Russia}

\begin{document}

\date{Accepted 2010..... Received 2010...; in original form 2010...}

\pagerange{\pageref{firstpage}--\pageref{lastpage}} \pubyear{2009}

\maketitle

\label{firstpage}

\begin{abstract}
We present and discuss new BVI CCD photometry in the field
of the globular cluster Arp~2, which is considered a member of the Sagittarius Dwarf Spheroidal
Galaxy. The main goal of this investigation
is to study of the statistics and spatial distribution of blue straggler
stars in the cluster. Blue stragglers are stars observed to be hotter and bluer than other stars with the same luminosity
in their environment. As such, they appear to be much younger than the rest of the stellar population. Two main
channels have been suggested to produce such stars: 
(1) collisions between stars in clusters or (2) mass transfer
between, or merger of, the components of primordial short-period binaries. The spatial distribution
of these stars inside a star cluster, compared with the distribution of stars
in different evolutionary stages,  can cast light on the most efficient production mechanism at work.
In the case of Arp~2, we found that blue straggler stars are significantly more concentrated than main sequence
stars, while they show the same degree of concentration as evolved stars (either red giants or horizontal branch stars). 
Since Arp~2 is not a very concentrated cluster, we suggest that this high central concentration
is an indication that blue stragglers are mostly primordial binary stars.
\end{abstract}


\section{Introduction}
Arp~2 is a globular cluster located at $l=8.54^{o}$, $b=-20.78^{o}$ ($\alpha=19^h~28^m~44^s$, 
$\delta=-30^{o}21^{\prime}~14^{\prime\prime}$, J2000.0).
It is commonly believed to have formed inside the Sgr dwarf spheroidal galaxy (Monaco et al. 2005),
and then released into the Milky Way through tidal interaction.
With a metallicity of [Fe/H]=-1.77 (Mottini et al. 2008), this cluster appears to be 3-4 Gyr younger
than the old globulars, but $\sim1-2$ Gyr older than  the youngest globulars
associated to Sgr (Carraro et al. 2007, Layden \& Sarajedini 2000).
The first photometric study of this cluster was performed by Buonanno et al. (1994).
The derived color-magnitude diagram (CMD) reveals an intriguing feature, namely
that the horizontal branch (HB) is located entirely blue-ward of the RR-Lyrae instability strip,
a fact that allowed the authors to assess its age through a differential
comparison with 47 Tuc and Ruprecht 106.
A secondary, prominent feature, which the authors do not comment on, is a group of stars
right above the turn-off (TO), which are probably the blue straggler stars (BSS)
population in Arp~2.
BSS are a normal stellar
population in clusters, since they are present in all of the properly observed Globular
Clusters (GC, Ferraro 2006; Ferraro et al. 2009, and reference therein).
Current scenarios for these stars in globulars are that either
they are binary system with significant mass exchange, or 
stellar mergers resulting from direct collisions between two or more stars 
(Davies et al. 2004; Knigge et al. 2009; Perets \& Frabrycky 2009).\\
In all these studies a proper assessment of the membership of BSSs through comparison
with Red Giant Branch (RGB) 
stars is routinely performed (Ferraro et al. 1993). The comparison of their 
cumulative radial distribution 
may hint to a possible common origin, specifically to confirm or deny whether
they belong to the same parent distribution.\\
Additionaly, the radial distribution of BSS in a star cluster is the most effective tool to 
understand their origin and which is the dominant production channel (Ferraro 2006).
BSS are routinely found - with the exception of Omega Cen and NGC~2419 (Dalessandro et al. 2008)- to be centrally concentrated.
Their radial profile then smooths down, while in the cluster periphery it shows again an
increase of the BSS contribution (Lanzoni et al. 2007, Dalessandro et al. 2008).\\

\noindent
In this paper we report on a new photometric data-set of Arp~2 obtained
with the goal of analyzing the BSS population in a young, relatively loose
GC, that is in an environment significantly different from a typically dense GC.
We anticipate here that our analysis reveals that BSS in Arp~2 do share the
same distribution of RGB and horizontal branch (HB) stars, and are more concentrated than 
main sequence (MS) stars.\\

\noindent
This paper is organized as follows. In Sect.~2 we illustrate
how we collected and analyzed our data, while in Sect.~3 we present
star counts in  Arp~2 to measure its size. The CMD of Arp~2 is presented and discussed
in Sect.~4, whereas the definition of BSS, and their statistics is discussed
in Sect.~6, which, also, comments on and summarizes our findings.

\begin{figure}
\includegraphics[width=\columnwidth]{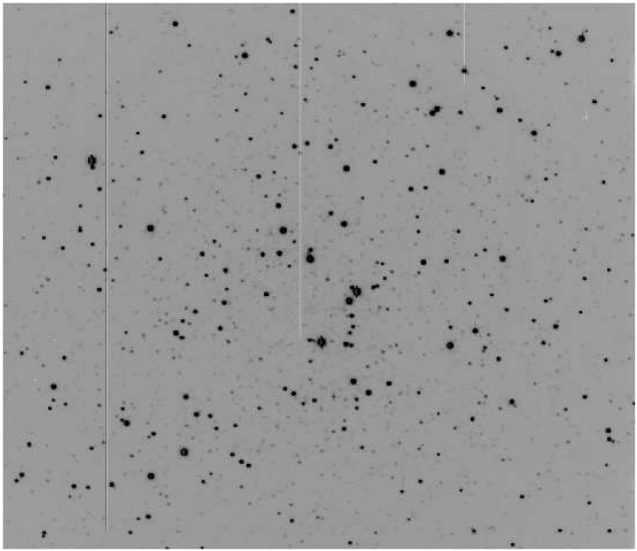}
\caption{A raw 900 sec image in the I filter of Arp~2. North is up, East to the left,
and the field of view if $\sim$ 4 arcmin on a side.}
\end{figure}

\begin{figure}
\includegraphics[width=\columnwidth]{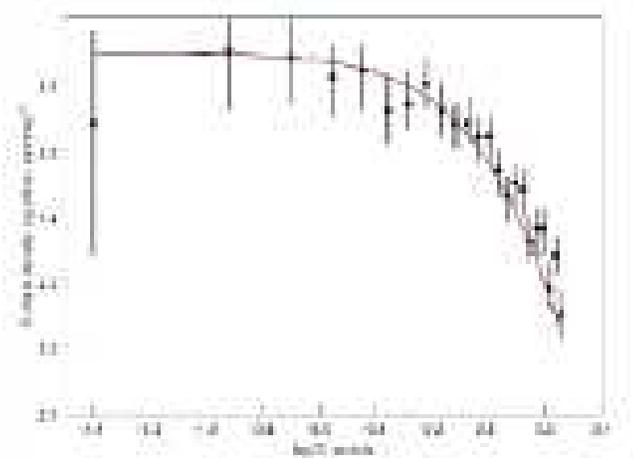}
\caption{Radial surface density profile as derived from our photometry. Over-imposed is a fit with a King profile, drawn adopting Harrin (1996) parameters}
\end{figure}

\section{Observations and Data Reduction}
CCD BVI images were acquired with the EFOSC2 camera mounted on the Nasmyth focus
of the ESO NTT
telescope on the night of August 8, 2008. The CCD is a 1030$\times$ 1038 array
with a scale of 0.24 arcsec, allowing to cover 4.1$\times$4.1 arcmin on the sky.
This allows us to cover a field centered on Arp~2 slightly larger than
the previous observations by Buonanno et al. (1994). We used multiple exposures
of 30 and 1200 secs for the B filter, and 30 and 900 both for V and I
filters.  As an illustration, an I filter, 900 sec, raw image is shown
in Fig.~1.
The night was photometric, and we calibrated our photometry
against the Landolt (1992) standard field Mark~A and PG~2213, observed
several times during the night. Arp~2 was observed during an observational
run focused on a different science when the principal target was not visible.
Data have been reduced in the standard way. Image preparation (trimming, bias and flatfield) 
was done using the
IRAF package, while photometry was extracted by using DAOPHOT and
ALLSTAR (Stetson 1987). 
We obtained a final catalog with 4580 entries having 2000.0 Equatorial coordinates,
and B, V and I magnitudes together with associated uncertainties.
These latter have been calculated following the prescriptions described
in Patat \& Carraro (2001).

\subsection{Complementary infrared data, astrometry and completeness}

Our optical catalogue was cross-correlated with 2MASS, which
resulted in a final catalog including $BVI$ and $JHK_{s}$ magnitudes. As a by product,
pixel (detector) coordinates were converted to RA and DEC for J2000.0 equinox, thus
providing 2MASS-based astrometry.\\

\noindent
Finally, completeness corrections were determined by running artificial star experiments
on the data. The data-set was divided in two regions, inner  (inside 1 arcmin) and outer
(beyond 1 arcmin), and
completeness was computed for these two different regions, which, due to the nature
of the object, are expected to be differently affected by star crowding (Carraro et al. 2007).
Basically, we created several artificial images by adding artificial stars
to the original frames. These stars were added at random positions, and had the same
color and luminosity distribution of the true sample. To avoid generating overcrowding,
in each experiment we added up to 30\% of the original number of stars. Depending on
the frame (short or deep exposures), 
between 1000-5000 stars were added. 
In this way we have estimated that the
completeness level of our photometry in the outer region is 100\% down to $V$ = 23.00, and better
than 50\% down to $V$ = 23.50. As for the inner region, we found that the completeness
is 100\% down to $V$ = 22.40, and better
than 50\% down to $V$ = 22.80.

\section{Star counts and cluster size}
We used our photometry to study the cluster stars radial distribution.
According to Harris (1996), Arp~2 has a concentration $c$=0.90,
an half-mass radius = 1.91 arcmin, and a core and tidal radius of 1.59 and 12.65
arcmin, respectively. Therefore we expect our photometry to cover just the
inner part of the cluster.
The radial density profile we constructed is shown in Fig.~2.
It has been derived following the  method described in Seleznev (1994).
This method employs numerical differentiation of the best mean-square 
polynomial fit for N(r), the number of stars in circles of radius r in
the plane of the sky.
The center of the cluster was taken at the detector coordinates (512,512) which
corresponds to $\alpha=19^h~28^m~44^s$, 
$\delta=-30^{o}21^{\prime}~14^{\prime\prime}$ (J2000.0). Vertical bars indicate the
profile error-bars derived assuming Poisson statistics.
In the same Fig.~2 we fit 
the profile with a  King (1962) model, adopting king parameters (core and tidal radius,
and center mass density) from Harris (1996) compilation.
The fit is good within the uncertainties, and therefore we conclude that Arp~2 
follows a King-like density profile. It 
decreases all the way to the edge of the field we covered,
slightly beyond the nominal half-mass radius.

\begin{figure}
\includegraphics[width=\columnwidth]{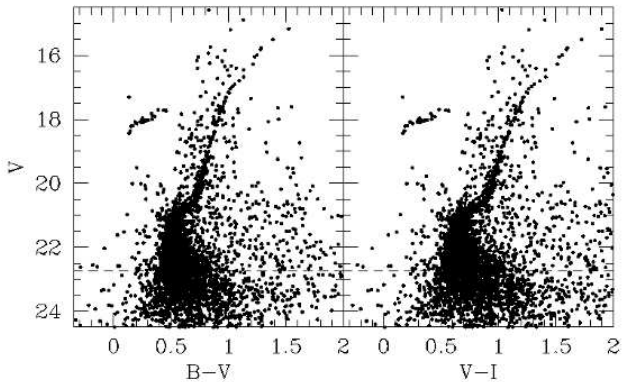}
\caption{CMD in the V vs B-V (left panel), and V vs V-I (right panel) for Arp~2.
The dashed lines indicate the 100$\%$ level of completeness. See Sect.~2.1}
\end{figure}
\begin{figure}

\includegraphics[width=\columnwidth]{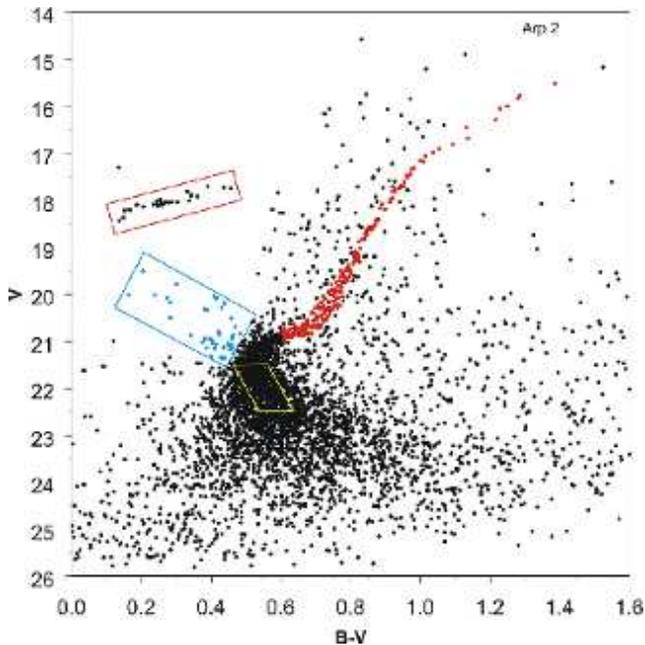}
\caption{Selection of RGB (red dots), HB (red circle), BSS (cyan circle) and MS (yellow
polygon) stars. These latter have been selected in a region of the MS where the completeness
is 100$\%$.}
\end{figure}

\section{Color Magnitude Diagrams}
The resulting color magnitude diagrams (CMDs) are shown in the two panels
of Fig.~3. In the left panel, the V versus B-V diagram is shown, while in the
right panel we present the V versus V-I.
The CMD on the left panel is absolutely identical to the one presented by Buonanno et al.(1994), apart from
the slightly different area coverage and the magnitude limit, which in our case
is about one magnitude fainter.
All the typical features of a globular cluster CMD are present. The MS,
RGB, HB -located entirely  blue-ward the
RR~Lyrae instability strip-,  and the Asimptotic Giant Branch (AGB).
An additional feature which has been overlooked in the past is the plume of blue stars
right above the turn off point (TO), which is quite common in globular clusters,
and it is composed of candidate blue straggler stars.
This plume is the target of our investigation. 
We look for BSS candidates following the commonly used criteria (Ahumada \& Lapasset
1995, 2007; Sandage 1953), which is illustrated in Fig.~4.
Together with BSS (blue box), we also indicated the location of 
HB (red box), RGB (red dots),  and a sample of MS stars (yellow box), which we are going to 
compare.
We counted 41 BSS candidates, and 28 HB stars.
They are listed in Table~1 and 2,
respectively, where we indicate stars' identification (our numbering, ID), 
equatorial coordinates for the 2000.0 equinox, magnitude V and color
B-V. These can also be useful for future spectro-scopic follow-up.\\
Besides, again following Fig~4, we counted 213 RGB and 517 MS stars, which we do not list here
for space reasons. 
As for MS stars, we stress that they have been extracted in a region of the MS
which is not affected by incompleteness.\\

\noindent
We stress that the numbers we reported have been computed using the classical
definition of BSS locus and assuming that contamination from field
stars is negligible, which seems to be the case, since the field
of view is very small (0.0044 squared degrees). However, we do not have at our disposal
a control field to verify this directly. Therefore, 
we investigated the amount of contamination by computing
synthetic CMDs of stars in the direction of Arp~2 assuming a Galactic model
which includes bulge, halo, thin and thick disks
(Girardi et a. 2005). We generated several CMDs by  varying the random
seed and added photometric errors as from Arp 2 photometry.
The results are shown in Fig.~5, where the left panel shows the Galaxy
CMD in the direction of Arp~2, and the right panel the CMD of Arp~2 as in Fig.~4.
In both panel we indicate the boxes used for selecting HB (red) and BSS (blue) stars.
A quick glance at this figure is sufficient to conclude that most of the contamination
affects the lower MS, and in general contaminating stars are redder than the typical
BSS colors. Some contamination is present in the RGB area but, provided the high
number of RGB stars in Arp~2, we do not expect their statistics to be significantly affected.

\begin{figure}
\includegraphics[width=\columnwidth]{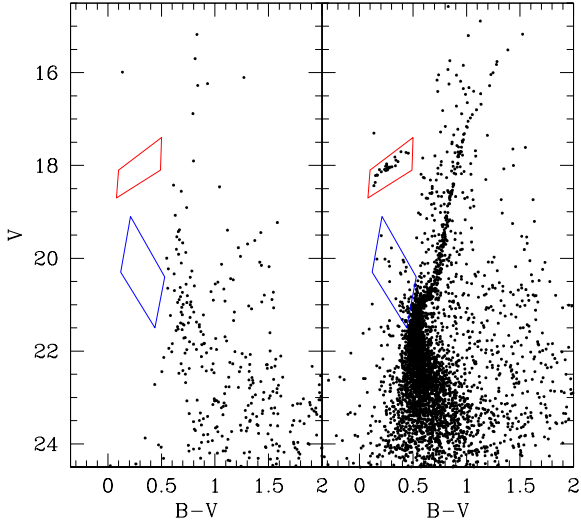}
\caption{Estimate of field star contamination in Arp~2. The left panel shows the Galaxy
CMD in Arp~2 direction, and the right panel the CMD of Arp~2 as in Fig.~4.}
\end{figure}

\begin{table}
\tabcolsep 0.1truecm
\caption{BSS candidates. See Fig.~4 for the selection}
\begin{tabular}{lcccc}
\hline
\noalign{\smallskip}
  ID   & $\alpha (2000.0)$ & $\delta(2000.0)$ & V & B-V\\
\hline
 140   & 292.1463587 & -30.3394516 & 19.777    & 0.299\\
 742   & 292.1539322 & -30.3574441 & 21.321    & 0.400\\
 909   & 292.1560467 & -30.3590792 & 20.831    & 0.482\\
 1086  & 292.1584272 & -30.3590217 & 21.173    & 0.453\\
 1119  & 292.1590122 & -30.3404261 & 20.071    & 0.274\\
 1463  & 292.1627267 & -30.3544101 & 20.191    & 0.437\\
 1575  & 292.1639108 & -30.3676803 & 21.347    & 0.438\\
 1649  & 292.1649792 & -30.3565927 & 21.040    & 0.448\\
 1942  & 292.1681176 & -30.3770277 & 20.871    & 0.340\\
 2257  & 292.1718385 & -30.3564573 & 20.680    & 0.469\\
 2308  & 292.1723293 & -30.3570290 & 20.282    & 0.430\\
 2324  & 292.1727334 & -30.3409102 & 20.315    & 0.382\\
 2341  & 292.1725939 & -30.3588043 & 21.154    & 0.410\\
 2433  & 292.1741541 & -30.3208680 & 20.853    & 0.423\\
 2531  & 292.1743077 & -30.3790334 & 21.371    & 0.431\\
 2568  & 292.1748233 & -30.3722961 & 21.268    & 0.462\\
 2750  & 292.1770090 & -30.3470938 & 21.341    & 0.430\\
 2826  & 292.1779222 & -30.3386991 & 21.040    & 0.403\\
 2910  & 292.1782726 & -30.3795017 & 21.024    & 0.374\\
 2913  & 292.1788245 & -30.3432700 & 20.935    & 0.460\\
 3146  & 292.1810024 & -30.3573719 & 20.947    & 0.396\\
 3163  & 292.1815525 & -30.3322570 & 20.343    & 0.388\\
 3266  & 292.1822050 & -30.3519277 & 21.131    & 0.386\\
 3351  & 292.1829014 & -30.3662331 & 20.172    & 0.281\\
 3362  & 292.1830748 & -30.3598515 & 20.039    & 0.408\\
 3533  & 292.1851738 & -30.3394198 & 20.981    & 0.402\\
 3633  & 292.1858386 & -30.3652012 & 20.082    & 0.414\\
 3637  & 292.1860246 & -30.3541686 & 21.078    & 0.458\\
 4014  & 292.1899276 & -30.3349827 & 20.018    & 0.163\\
 4140  & 292.1907422 & -30.3659681 & 20.510    & 0.458\\
 4346  & 292.1932713 & -30.3685683 & 20.874    & 0.462\\
 4420  & 292.1946120 & -30.3279961 & 21.127    & 0.435\\
 4635  & 292.1967567 & -30.3540156 & 19.512    & 0.203\\
 4875  & 292.1993316 & -30.3591077 & 21.066    & 0.413\\
 5059  & 292.2014691 & -30.3478313 & 22.921    & 0.668\\
 5174  & 292.2029316 & -30.3524965 & 20.789    & 0.421\\
 5239  & 292.2037376 & -30.3534943 & 20.504    & 0.282\\
 5354  & 292.2051005 & -30.3629483 & 21.345    & 0.428\\
 5611  & 292.2091052 & -30.3263457 & 20.941    & 0.382\\
 5612  & 292.2085462 & -30.3696267 & 21.149    & 0.460\\
 6087  & 292.2157226 & -30.3813623 & 20.019    & 0.238\\
\hline
\end{tabular}
\end{table}

\begin{table}
\tabcolsep 0.1truecm
\caption{HB candidates. See Fig.~4 for the selection}
\begin{tabular}{lcccc}
\hline
\noalign{\smallskip}
  ID   & $\alpha (2000.0)$ & $\delta(2000.0)$ & V & B-V\\
\hline 
 125  & 292.1456946 & -30.3733031 & 17.819 &  0.331\\ 
 370  & 292.1493758 & -30.3477929 & 18.091 &  0.260\\ 
 1127 & 292.1586924 & -30.3658908 & 18.069 &  0.254\\ 
 1549 & 292.1641508 & -30.3352455 & 17.735 &  0.456\\ 
 1996 & 292.1690971 & -30.3519957 & 18.139 &  0.204\\ 
 2040 & 292.1691049 & -30.3828881 & 18.207 &  0.162\\ 
 2086 & 292.1701074 & -30.3484238 & 17.966 &  0.253\\ 
 2503 & 292.1746582 & -30.3377476 & 18.087 &  0.256\\ 
 2608 & 292.1753085 & -30.3659882 & 18.219 &  0.150\\ 
 3036 & 292.1802420 & -30.3302122 & 18.007 &  0.336\\ 
 3038 & 292.1802864 & -30.3280083 & 18.367 &  0.147\\ 
 3129 & 292.1807069 & -30.3656050 & 18.089 &  0.186\\ 
 3204 & 292.1815674 & -30.3589202 & 18.038 &  0.271\\ 
 3555 & 292.1848119 & -30.3832435 & 18.087 &  0.234\\ 
 3682 & 292.1864725 & -30.3582211 & 17.996 &  0.253\\ 
 3772 & 292.1871954 & -30.3651926 & 17.886 &  0.341\\ 
 3990 & 292.1894482 & -30.3534246 & 17.703 &  0.388\\ 
 4040 & 292.1898984 & -30.3533154 & 17.917 &  0.360\\ 
 4911 & 292.1998576 & -30.3476994 & 18.097 &  0.258\\ 
 4932 & 292.2002557 & -30.3352233 & 17.720 &  0.435\\ 
 4940 & 292.1997888 & -30.3748761 & 18.209 &  0.168\\ 
 5048 & 292.2011579 & -30.3552338 & 18.054 &  0.277\\ 
 5101 & 292.2019892 & -30.3501858 & 18.437 &  0.135\\ 
 5165 & 292.2033021 & -30.3208588 & 18.108 &  0.245\\ 
 5184 & 292.2029381 & -30.3599449 & 18.018 &  0.243\\ 
 5244 & 292.2038283 & -30.3544071 & 18.028 &  0.304\\ 
 5646 & 292.2087883 & -30.3863411 & 18.051 &  0.266\\ 
 5773 & 292.2115368 & -30.3289674 & 18.037 &  0.287\\  
\hline
\end{tabular}
\end{table}

\begin{figure}
\includegraphics[width=\columnwidth]{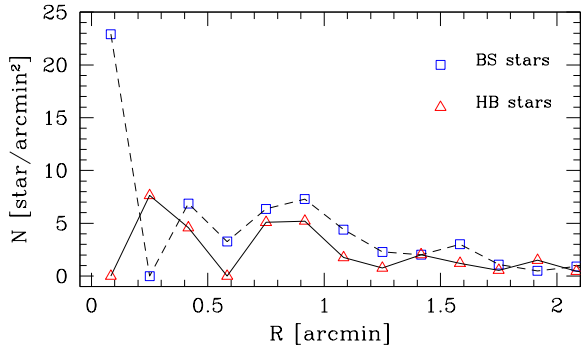}
\caption{Radial surface density profile of HB (red triangles) and BSS (blue squares) stars i
n the field of Arp~2}
\end{figure}

\begin{figure}
\includegraphics[width=\columnwidth]{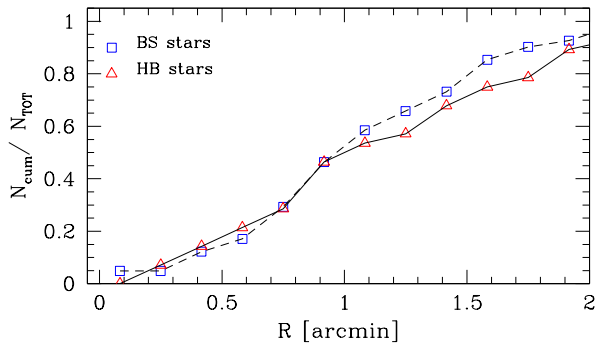}
\caption{Fractional radial cumulative distribution of BSS and HB stars}
\end{figure}

\section{Analysis of the Blue Straggler Stars' population}
To properly assess the probable membership of BSS, it is necessary to measure their
radial velocity or their proper motion (Mathieu \& Geller 2009; Liu et al. 2008). 
Since we are relying only on
photometry, we will keep considering our BSS as candidates, based on their
location in the CMD.
To get more insight on their properties and origin, we started by considering their
surface distribution, and compared it with the  surface distribution
of HB stars. 
This is illustrated in Fig~6, which shows radial density profile
of the two populations. Here we plot the number
of stars per squared arcmin as computed in concentric bins 10 arcsecs wide.
Except for the difference in the inner side of the cluster,  the two profiles are identical.
We stress that the difference in the very central bin, although real,
is exagerated by the small number statistics (2 BSS and 0 HB stars).
In this respect Arp~2 seems more similar to old open clusters, like M~67 
(Mathieu \& Gheller 2009), than to genuine globulars.
The higher concentration of BSS in the cluster internal region with respect to evolved stars
we find is not new,
and has been found already in several other globulars (Dalessandro et al 2008).
At odds with what is found in other globulars, we do not see any outward increase
of the BSS population, most probably because we are not sampling the cluster
outskirts (see Section~3)in this study .

\section{Discussion and Conclusion}
\noindent
To better quantify the relationship between BSS and the other stars 
we make use of Kolmogorov-Smirnov (KS) statistics,
in its one (1D) and two (2D) dimensional flavors.\\
The 2D distributions of BSS with respect to other star  samples
in different evolutionary phases (HB, RGB or MS stars) 
were statistically compared with the
2D generalization of 1D KS test
described in Press et al. (1997) and Fasano \& Franceschini (1987). 
The same method was employed 
in Pancino et al. (2003) for comparison of the 2D
distributions of RGB stars with different metallicity in the 
multi-population globular
cluster $\omega$ Cen.\\

\noindent
This test gives the probability P that two distributions are extracted
from the same parent distribution. Small values of  P would show that the two
samples are significantly different. Formulas as given in Press et al. (1997)
are accurate enough when 

\begin{equation}
N = N_{1} \cdot N_{2}/( N_{1}+N_{2}) > 20
\end{equation}

\noindent
and when the
indicated probability P is less (more significant than) 0.20 or so.
In the above equation N1 and N2 are the two populations under comparison.
When P is larger than 0.20, its value may not be accurate, but the implication that
the two data sets are not significantly different is certainly correct.
We summarize our results in Table~3 and 4, and graphically in the series of Figs.~7 to 9.

\begin{figure}
\includegraphics[width=\columnwidth]{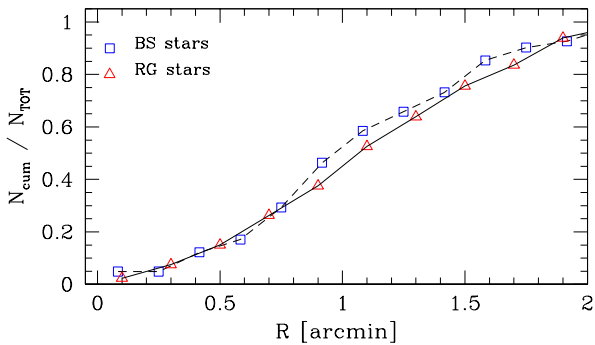}
\caption{Fractional radial cumulative distribution of BSS and RGB stars}
\end{figure}

\begin{figure}
\includegraphics[width=\columnwidth]{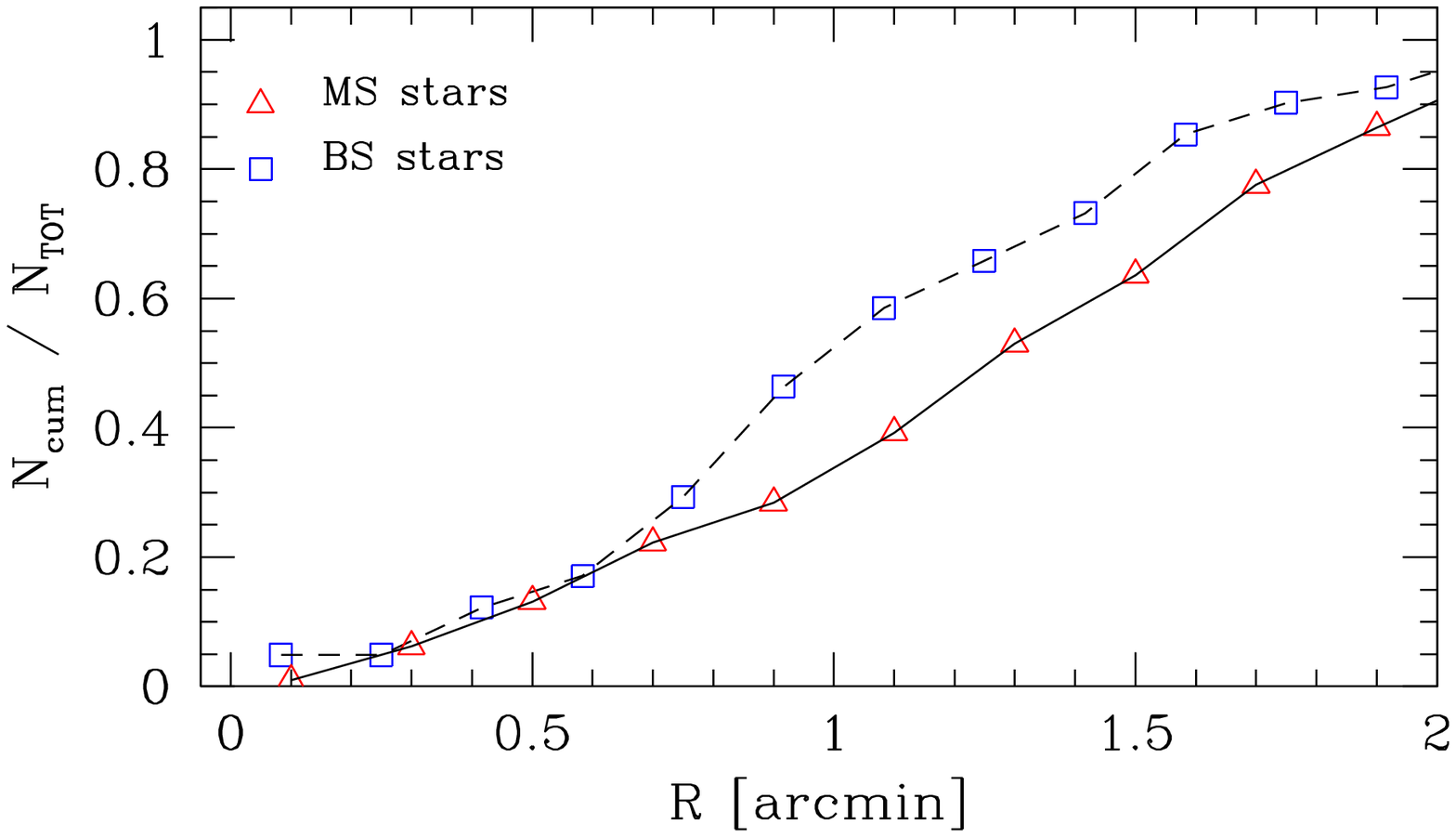}
\caption{Fractional radial cumulative distribution of BSS and MS stars}
\end{figure}

\begin{table}
\tabcolsep 0.1truecm
\caption{Results of the KS statistics in 1D for the BSS. }
\begin{tabular}{lccc}
\hline
\noalign{\smallskip}
 Population & N & P & Reference Figure\\
\hline
  HB & 17 &  0.83 & Fig~7\\
 RGB & 35 &  0.61 & Fig~8\\
  MS & 38 &  0.07 & Fig~9\\
\hline 
\end{tabular}
\end{table}

\begin{table}
\tabcolsep 0.1truecm
\caption{Results of the KS statistics in 2D for the BSS. }
\begin{tabular}{lcc}
\hline
\noalign{\smallskip}
 Population & N & P \\
\hline
  HB & 17 & 0.51\\
 RGB & 35 & 0.43\\
  MS & 38 & 0.23\\
\hline 
\end{tabular}
\end{table}

\noindent
Looking at the results listed in the two tables and illustrated in the corresponding figures, 
we can provide the following
considerations.\\

\noindent
First, due to the larger number, the statistics (N) improves passing from HB to RGB stars,
and from RGB to MS stars.\\

\noindent
Second, probabilities P for 2D KS test are smaller than for 1D test in the case of HB and RGB stars. 
We face the opposite situation as for MS stars. 
We can explain it by taking into account that 1D distribution
is only a function of the distance from the distribution
center, while 2D distributions do contain angle information.
In the case of HB and RGB stars radial distributions are
very close to BSS, while azimuthal distributions
have some differences, and this results in larger probabilities
in 1D testing. In the case of MS stars radial distribution is
very different from BSS distribution, and probability
in 1D test is small. Apparently, azimuthal distributions in this
case are more similar, which implies larger probability in
2D test.\\

\noindent
Third, and more interesting, the tests provide comparable results for HB and RGB stars.
This implies that BSS, HB and RGB stars are on the overall 
distributed in the same way, which in turn  means that these populations
are not significantly different. 
We caution however that high probability in KS test is a necessary, but not sufficient
condition 
to prove that these populations
have the same parent distribution (see e.g. Press et al., 1997).\\

\noindent
At odds with HB and RGB, MS stars show a different distribution when compared to  BSS.
Namely, BSS are significantly more concentrated to the cluster center than normal MS stars.
All-together this suggests that most probably BSS are primordial binary systems which,
because of their total mass and the relatively loose environment of Arp~2, sank  toward the center
and survived as binary system.\\

\noindent
Our results also imply that present-day RGB and HB stars are, on the overall, more concentrated than
actual MS stars. This does not mean that actual HB stars and upper RGB stars are more massive
than actual MS stars, which can be the case for RGB stars in earlier stages of evolution. 
This simply reflects the fact that actual upper RGB and HB stars follow the spatial distribution
of their -originally more massive- MS progenitors. The relaxation time for Arp~2 is $\sim$5 Gyr (Harris 1996),
specifically much shorter than the cluster age, and therefore mass segregation already occured in the past,
when the progenitors of present-day upper RGB and HB stars were in the MS, and were more massive - and therefore
more segregated - than the
present-day MS stars.\\

\noindent
Our findings confirm very recent results on the BSS distribution and nature in GCs
(Ferraro et al. 2009).\\
The suggestions (Knigge et al. 2009) that most BSS in globulars
have a binary origin, even in the environment of the densest globulars,
is not longer of general validity.  
In fact, whether BSS are primordial close binaries or merger remnants resulting from direct collision between 
two stars, depends strongly
on the environment.
As recently discovered by Ferraro et al. (2009),
the existence of double BSS sequences in M30, a famous {\it core collapse} globular cluster,
confirms that BSS nature depends strongly on the environment, and the dynamical state of the parent
cluster. \\
Loose systems, such as open clusters, do indeed  show a high primordial binary fraction among
their BSS (Mathieu \& Gheller 2009), while very dense systems harbour both primordial
binaries and merger products, in proportions which vary from cluster to cluster.\\

\noindent
In this context, Arp~2 is  a relatively low concentration globular, that follows this trend, 
which will
be hopefully confirmed by future spectroscopic observations.

\section*{Acknowledgments}
GC expresses his gratitude to O. Hainaut and G. Lo Curto for assistance
during the observations, and Y. Momany for long useful conversations on
blue stragglers. We express special thanks to the referee, Robert Rood,
for the careful reading and useful suggestions he provided. 
Finally, we expresse our gratitude to Sandy Strunk for reading
carefully the manuscript and improving the language.
In preparation of this paper, we made use of the NASA Astrophysics 
Data System and the ASTRO-PH e-print server. This work  made extensive use of
the SIMBAD database, operated at the CDS, Strasbourg, France.

\end{document}